\documentstyle[a4,12pt]{article}
\begin{document}
\pagestyle{empty}
\begin{titlepage}
\rightline{IC/96/286}
\rightline{UTS-DFT-96-04}
\vspace{2.0 truecm}
\begin{center}
\begin{Large}
{\bf $Z^\prime$ effects and anomalous gauge couplings} \\ [0.4cm]
{\bf at LC with polarization}
\end{Large}

\vspace{2.0cm}

{\large  A. A. Pankov\hskip 2pt\footnote{Permanent address: Gomel
Polytechnical Institute, Gomel, 246746 Belarus.\\ 
E-mail PANKOV@GPI.GOMEL.BY}
}\\[0.3cm]
International Centre for Theoretical Physics, Trieste, Italy\\

\vspace{5mm}

{\large  N. Paver\hskip 2pt\footnote{Also supported by the Italian
Ministry of University, Scientific Research and Technology (MURST).}
}\\[0.3cm]
Dipartimento di Fisica Teorica, Universit\`{a} di Trieste, 34100
Trieste, Italy\\
Istituto Nazionale di Fisica Nucleare, Sezione di Trieste, 34127 Trieste,
Italy

\vspace{5mm}

{\large  C. Verzegnassi} 
\\[0.3cm]
Dipartimento di Fisica, Universit\`{a} di Lecce, 73100 Lecce, Italy\\
Istituto Nazionale di Fisica Nucleare, Sezione di Lecce, 73100 Lecce,
Italy\\
\end{center}

\vspace{2.0cm}

\begin{abstract}
\noindent 
We show that the availability of longitudinally polarized electron beams at 
a $500\hskip 2pt GeV$ Linear Collider would allow, from an analysis of 
the reaction $e^+e^-\to W^+W^-$, to set stringent bounds on the couplings of 
a $Z^\prime$ of the most general type. In addition, to some extent, it would be 
possible to disentangle observable effects of the $Z^\prime$ from analogous 
ones due to competitor models with anomalous trilinear gauge couplings. 
\vspace*{3.0mm}

\noindent  
\end{abstract}
\end{titlepage}

\pagestyle{plain}
\setlength{\baselineskip}{1.3\baselineskip}
\section{Introduction}
It has been recently suggested \cite{verze1} that theoretical models with one 
extra $Z\equiv Z^\prime$ whose couplings to quarks and leptons are not of the 
`conventional' type would be perfectly consistent with all the available 
experimental information from either LEP1 \cite{lep1} and SLD \cite{sld} 
or CDF \cite{cdf} data. Starting from this observation, a detailed analysis 
has been performed of the detectability in the final two-fermion channels at 
LEP2 of a $Z^\prime$ whose fermion couplings are arbitrary (but still family 
independent) \cite{verze2}. Also, in \cite{verze2} the problem of 
distinguishing this model from competitor ones (in particular, from a model 
with anomalous gauge couplings) has been studied.
\par 
The final two-fermion channel is not the only one where virtual effects 
generated by a $Z^\prime$ can manifest themselves. The usefulness of the final 
$W^+W^-$ channel in $e^+e^-$ annihilation to obtain improved information on 
some theoretical properties of such models, has already been stressed in 
previous papers in the specific case of longitudinally 
polarized beams for models of `conventional' type (e.g., $E_6$, $LR$, 
{\it etc.}), showing that the role of polarization in these cases would be 
essential \cite{sasha}.
\par
The effects of a $Z^\prime$ of `unconventional' type in the $W^+W^-$ channel 
have also been considered, and compared with those of models with anomalous 
gauge couplings in Ref.~\cite{verze3}. In particular, in \cite{verze3} it was 
shown that the benchmark of the model with a $Z^\prime$ would be the existence 
of a peculiar connection between certain effects observed in the $W^+W^-$ 
channel and other effects observed in the final lepton-antilepton channel.  
\par
The aim of this paper is that of considering whether the search for 
{\it indirect} effects 
of a `unconventional' $Z^\prime$ in the $W^+W^-$ channel would benefit from 
the availability of longitudinal polarization of initial beams, as it is the 
case for the `conventional' situation. We shall show in the next Sect. 2 that 
this is indeed the case, i.e., that in the parameter space the expected 
experimental sensitivity in the polarized processes is by far better than in 
the unpolarized case. For what concerns the differentiation from other 
sources of nonstandard effects, in particular those with anomalous 
gauge couplings, we shall also 
show in Sect. 3 that the characteristic feature of such a $Z^\prime$ would be 
the existence of certain peculiar properties of different observables, all 
pertaining to the final $W^+W^-$ channel.
\par
All our discussions assume that longitudinal lepton polarization will be 
available in the considered examples. In practice, this would be feasible at 
the future planned $500\hskip 2pt GeV$ linear Collider (LC). Our conclusions  
for the specific cases that we consider, as summarized in  
Sect. 4,  will therefore be strongly in favour of polarization. An 
Appendix will be devoted to the derivation of several expressions for the 
relevant experimental observables. 

\section{Derivation of the constraints for general $Z^\prime$ parameters}  
The starting point of our analysis will be the expression of the invariant 
amplitude for the process
\begin{equation} e^++e^-\to W^++W^-.\label{proc}\end{equation}
In Born approximation, this can be written as a sum of a $t$-channel and of an 
$s$-channel component. In the Standard Model (SM) case, the latter will be 
written as follows: 
\begin{equation}
{\cal M}_s^{(\lambda)}=\left(-\frac{1}{s}
+\frac{\cot\theta_W (v-2\lambda a)}{s-M^2_Z}\right)
\times{\cal G}^{(\lambda)}(s,\theta),\label{amplis}\end{equation} 
where $s$ and $\theta$ are the total c.m. squared energy and $W^-$ production 
angle; $v=(T_{3,e}-2Q_e\hskip 2pt s_W^2)/2s_Wc_W$ 
and $a=T_{3,e}/2s_Wc_W$ with $T_{3,e}=-1/2$ and $s_W=\sin\theta_W$, 
$c_W=\cos\theta_W$ ($\theta_W$ is the conventional electroweak mixing angle); 
$\lambda$ denotes the 
electron helicity ($\lambda=\pm1/2$ for right/left-handed electrons); 
finally, ${\cal G}^{(\lambda)}(s,\theta)$ is a kinematical coefficient,  
depending also on the final $W$'s helicities. For simplicity we omit its 
explicit form, which is not essential for our discussion here, and can be 
either found in the literature \cite{gounaris} or easily derived from the 
entries of Tab.~1 in the subsequent Sect. 3, which also shows the form of the 
$t$-channel neutrino exchange.  
Note that, at this stage, we are writing an `effective' Born approximation that 
contains both the physical $Z$ couplings and the physical $Z$ mass. We shall 
systematically ignore extra contributions at one loop. In fact, our purpose 
is that of evaluating {\it deviations} from the SM expressions due to one 
extra $Z$. In this spirit, we shall also consider the $Z^\prime$ contribution 
using an `effective' Born approximation with physical $Z^\prime$ couplings and 
mass. The more rigorous one-loop treatment would require also the calculation 
of the, potentially dangerous, QED radiation effects whose study has not yet 
been performed, to our knowledge, for polarized beams at the LC. We shall 
assume in the sequel that the results of a rigorous treatment reproduce those 
of an effective approximation without QED after a suitable, apparatus 
dependent, calculation as it is the case for the unpolarized case. Then, for 
the evaluation of the {\it deviations} due to the $Z^\prime$, the residual 
purely electroweak one-loop contributions will be safely neglected.
\par
Working in this framework, the effective expression of the invariant amplitude 
after addition of one extra $Z$ will be written as: 
\begin{equation}{\cal M}_s^{(\lambda)}=\left(-\frac{1}{s}
+\frac{g_{WWZ_1} (v_1-2\lambda a_1)}{s-M_{Z_1}^2}
+\frac{g_{WWZ_2} (v_2-2\lambda a_2)}{s-M_{Z_2}^2}\right)
\times{\cal G}^{(\lambda)}(s,\theta).\label{amplis1}\end{equation}
In Eq.~(\ref{amplis1}), we have retained two possible sources of effects from, 
respectively, the `light' and the `heavy' neutral gauge bosons $Z_1$ and 
$Z_2$. In general, in models with two neutral gauge bosons, the `light' $Z$ 
is formally not identical to the SM $Z$ and, 
within the accuracy of experimental data relevant to measurements of the 
standard $Z$ parameters, it potentially can have mass and couplings different 
from the SM prediction. Such modifications of the $Z$ couplings, reflecting the 
presence of the additional extra $Z^\prime$, can be induced, e.g., through the 
mechanism of $Z-Z^\prime$ mixing. To account for this fact, the `light' $Z$ 
is now denoted as $Z_1$, and the same convention applies to its couplings 
$v_1$, $a_1$ and $g_{WWZ_1}$. The second effect is due to the actual extra 
heavy $Z$ exchange diagram, and will be treated by denoting the physical 
heavy $Z$ as $Z_2$ and its physical couplings by analogous 
notations.\footnote{In Eq.~(\ref{amplis1}), the couplings to 
$W^+W^-$ of both $Z_1$ and $Z_2$ have been tacitly assumed of the usual 
Yang-Mills form.}
\par 
It turns out that it is convenient to rewrite Eq.~(\ref{amplis1}) in the 
following form:  
\begin{equation} 
{\cal M}_s^{(\lambda)}=\left(-\frac{g_{WW\gamma}}{s}
+\frac{g_{WWZ}(v-2\lambda a)}{s-M^2_Z}\right)
\times{\cal G}^{(\lambda)}(s,\theta),\label{amplis2}\end{equation}
where the `effective' gauge boson couplings $g_{WW\gamma}$ and $g_{WWZ}$ 
are defined as: 
\begin{equation} 
g_{WW\gamma}= 1+\Delta_\gamma= 
1+\Delta_\gamma(Z_1)+\Delta_\gamma(Z_2),\label{deltagamma}\end{equation}
\begin{equation}
g_{WWZ}=\cot\theta_W+\Delta_Z= 1+\Delta_Z(Z_1)+\Delta_Z(Z_2),
\label{deltaz}\end{equation}
with
\begin{equation}\Delta_\gamma(Z_1)=v\hskip 2pt \cot\theta_W\hskip 2pt 
\left(\frac{\Delta a}{a}-\frac{\Delta v}{v}\right)\hskip 1pt
\left(1+\Delta\chi\right)\hskip 1pt\chi;\ \   
\Delta_\gamma(Z_2)=v\hskip 2pt g_{WWZ_2}\hskip 2pt
\left(\frac{a_2}{a}-\frac{v_2}{v}\right)\hskip 2pt \chi_2,
\label{coupl1}
\end{equation}
\begin{equation}
\Delta_Z(Z_1)=\Delta g_{WWZ}+\cot\theta_W\hskip 1pt\left(\frac{\Delta a}{a}
+\Delta\chi\right);\qquad \Delta_Z(Z_2)=
g_{WWZ_2}\hskip 2pt \frac{a_2}{a}\hskip 2pt \frac{\chi_2}
{\chi}.\label{coupl2}\end{equation}
In Eqs.~(\ref{coupl1}) and (\ref{coupl2}) we have introduced the deviations 
of the fermionic and trilinear bosonic couplings $\Delta v=v_1-v$, 
$\Delta a=a_1-a$ and $\Delta g_{WWZ}=g_{WWZ_1}-\cot\theta_W$, and the 
neutral vector boson propagators (neglecting their widths):
\begin{equation}
\chi (s)=\frac{s}{s-M^2_Z};\qquad \chi_2 (s)=\frac{s}{s-M^2_{Z_2}};
\qquad \Delta\chi (s)=-\frac{2M_Z\Delta M}{s-M^2_Z},\label{chi}
\end{equation}
where $\Delta M=M_Z-M_{Z_1}$ is the $Z$-$Z_1$ mass-shift. 
\par  
It should be stressed that, not referring to specific models, the 
parametrization (\ref{amplis2})-(\ref{deltaz}) is both general and useful for 
phenomenological purposes, in particular to compare different sources 
of nonstandard effects contributing finite deviations (\ref{coupl1}) and 
(\ref{coupl2}) to the SM predictions.
\par 
Concerning the $Z_1$ couplings to electrons, present constraints from
experimental data indicate that their values should be rather close to the SM 
values $v$ and $a$ listed above.
Indeed, results for the effective leptonic vector and axial-vector couplings
derived from the combined LEP data give $v/v_{SM}=1.0008\pm 0.0235$ and 
$a/a_{SM}=0.9902\pm 0.0006$ \cite{lep_recent}, and the analysis 
of the $\rho$-parameter suggests an upper limit on $\Delta M$ of the order of 
$150-200\hskip 1pt MeV$ \cite{altarelli, babu}. Thus, the deviations 
$\Delta v$ and $\Delta a$ should be small numbers, to be treated 
as a perturbation to the SM results, and the same is true for the mass-shift 
$\Delta M/M_Z$, with $\Delta M>0$ if this is due to ${Z-Z^\prime}$ mixing.
\par 
As we are interested here in the sensitivity of process (\ref{proc}) to 
general features of virtual, `indirect', $Z^\prime$ effects and in their 
comparison with analogous effects of anomalous gauge boson couplings, we 
do not consider modifications of the $t$-channel amplitude. In principle, 
such effects can arise in specific models due to the presence of 
new heavy fermions, and their inclusion in the general parametrization of 
the amplitude deviations from the SM would require a much more complicated 
analysis, taking into account three different sources of nonstandard 
effects at the same time ($Z^\prime$, lepton mixing and anomalous couplings).  
Although perhaps possible, this is beyond the purpose of this 
paper.\footnote{For an attempt of separating heavy-lepton mixing effects from 
$Z^\prime$ ones see, e.g., Ref.~\cite{babich}.}
\par We now focus on the effects of the heavy $Z$ 
on polarized observables. Although this is not necessarily a unique 
choice, we mostly consider for a first investigation the case of only polarized 
electron beams. The general expression for the cross section of process 
(\ref{proc}) with longitudinally polarized electron and positron beams can be 
expressed as
\begin{equation}
\frac{d\sigma}{d\cos\theta}=\frac{1}{4}\left[\left(1+P_L\right)
\left(1-{\bar P}_L\right)\frac{d\sigma^+}{d\cos\theta}+
\left(1-P_L\right)\left(1+{\bar P}_L\right)\frac{d\sigma^-}{d\cos\theta}
\right], \label{longi}\end{equation}
where $P_L$ and ${\bar P}_L$ are the actual degrees of electron 
and positron longitudinal polarization, respectively, and 
$\sigma^\pm$ are the cross sections for purely right-handed and left-handed 
electrons. From Eq.~(\ref{longi}), the cross section for polarized 
electrons and unpolarized positrons corresponds to ${\bar P}_L=0$.
The polarized cross sections can be generally decomposed as follows:
\begin{equation}
\frac{d\sigma^\pm}{d\hskip 1pt \cos\theta}=
\frac{\pi\alpha^2_{em}\beta_W}{2s}\hskip 2pt\sum_i 
F_i^\pm\hskip 1pt {\cal O}_i(s,\cos\theta),\label{xsection}\end{equation}
where: $\beta_W=\sqrt{1-4M_W^2/s}=2p/\sqrt s$, with $p=\vert\vec p\vert$ the 
CM momentum of the $W$; $F_i^\pm$ are combinations of couplings involving in 
particular the deviations from the SM couplings, e.g., of the kind previously 
introduced; ${\cal O}_i$ are functions of the kinematical variables. To make 
the paper self-contained, we list the explicit expressions of the relevant 
$F_i^\pm$ and ${\cal O}_i$ in the Appendix.  
\par 
In practice, we shall denote by $\sigma^L$ and $\sigma^R$ the 
cross sections corresponding, in Eq.~(\ref{longi}), to the values $P_L=-0.9$ 
and $P_L=0.9$, respectively. Such degrees of longitudinal polarization should 
be realistically obtainable at the LC \cite{prescott}. 
\par
Our analysis proceeds in this way. Following the suggestions of 
previous dedicated searches \cite{pankov}, the sensitivity of 
$\sigma^L$ and $\sigma^R$ to $\Delta_\gamma$ and $\Delta_Z$ is assessed 
numerically by dividing the angular range $\vert\cos\theta\vert\leq 0.98$ 
into 10 equal `bins', and defining a $\chi^2$ function in terms of the expected 
number of events $N(i)$ in each bin:
\begin{equation}
\chi^{2}=\sum^{bins}_i\left[\frac{N_{SM}(i)-N(i)}
{\delta N_{SM}(i)}\right]^2,\label{chi2}\end{equation}
where the uncertainty on the number of events $\delta N_{SM}(i)$ combines 
both statistical and systematic errors as
\begin{equation} \delta N_{SM}(i)=
\sqrt{N_{SM}(i)+\left(\delta_{syst}N_{SM}(i)\right)^2},
\label{deltan} \end{equation}
(we assume $\delta_{syst}=2\%$).
In Eq.~(\ref{chi2}), $N(i)=L_{int}\sigma_i\varepsilon_W$ with $L_{int}$ the 
time-integrated luminosity, and ($z=\cos\theta$):
\begin{equation}
\sigma_i=\sigma(z_i,z_{i+1})=
\int \limits_{z_i}^{z_{i+1}}\left({d\sigma}\over{dz}\right)dz,
\label{sigmai}\end{equation}
Finally, $\varepsilon_W$ is the efficiency for $W^+W^-$ reconstruction, 
for which we take the channel of lepton pairs ($e\nu+\mu\nu$) plus two 
hadronic jets, giving $\varepsilon_W\simeq 0.3$ from the relevant branching 
ratios. An analogous procedure is followed to evaluate $N_{SM}(i)$.
\par 
As a criterion to derive the constraints on the coupling constants in the 
case where no deviations from the SM were observed, we impose that 
$\chi^2\leq\chi^2_{crit}$, where $\chi^2_{crit}$ is a number that specifies 
the chosen confidence level. With two independent parameters in 
Eqs.~(\ref{deltagamma}) and (\ref{deltaz}), the $95\%$ CL is 
obtained by choosing $\chi^2_{crit}=6$ \cite{pdg}.
\par 
From the numerical procedure outlined above, we obtain the allowed bands 
for $\Delta_\gamma$ and $\Delta_Z$ determined by the polarized cross sections 
$\sigma^R$ and $\sigma^L$ (as well as $\sigma^{unpol}$) depicted in Fig.~1, 
where $L_{int}=50\hskip 2pt fb^{-1}$ has been assumed.
\par
One can see from inspection of Fig.~1 that the role of polarization is 
essential 
in order to set meaningful finite bounds. Indeed, contrary to the unpolarized 
case, which evidently by itself does not provide any finite region for 
$\Delta_\gamma$ and $\Delta_Z$ (unless one of the two parameters is fixed by 
some further assumption), from the combined and {\it intersecting} 
bands relative to $\sigma^L$ and $\sigma^R$ one can derive the following 
95\% CL allowed ranges
\begin{equation}\begin{array}{c}
-0.002 <  \Delta_\gamma  < 0.002 \\
-0.004 <  \Delta_Z  < 0.004.\end{array}\label{del1}\end{equation}
\par 
Reflecting the generality of the parametrization 
(\ref{amplis2})-(\ref{deltaz}), Eq.~(\ref{del1}) represents the most general 
constraint that would 
be derivable at the LC for a `unconventional' $Z^\prime$ with polarized 
electron beams. It should be stressed that the constraint is completely model 
independent. To have a feeling of how polarization works in more specific 
cases, we consider as an application the familiar situation of an extra $Z$ of 
extended gauge origin, in particular generated by a previous $E_6$ 
symmetry \cite{rizzo}. Denoting by $\phi$ the $Z$-$Z^\prime$ mixing angle 
defined by following the conventional prescriptions, Eqs.~(\ref{coupl1}) 
and (\ref{coupl2}) would read now:
\begin{equation}
\Delta_\gamma=v\cot\theta_W \phi 
\left(\frac{a^\prime}{a}-\frac{v^\prime}{v}\right)\hskip 1pt 
\left(1-\frac{\chi_2}{\chi}+\Delta\chi\right)\chi,\label{ddeltag}\end{equation}
\begin{equation}
\Delta_Z=\cot\theta_W\hskip 1pt\left[\phi\hskip 1pt\frac{a^\prime}{a}\hskip 1pt
\left(1-\frac{\chi_2}{\chi}\right)+\Delta\chi\right],\label{ddeltaz}
\end{equation}
with $v^\prime$ and $a^\prime$ fixed by the specific model and, in general 
\begin{equation}
\tan^2\phi=\frac{M_Z^2-M_{Z_1}^2}{M_{Z_2}^2-M_Z^2}
\simeq\frac{2M_Z\Delta M}{M_{Z_2}^2}.\label{phi}\end{equation}
 
Consequently, in the ($\Delta_\gamma,\Delta_Z$) plane of Fig.~1 each model is 
now represented, in linear approximation in $\phi$, by a line of 
equation:\footnote{Although present in general, $\Delta\chi$ can be safely 
neglected as being quadratic in the small mixing angle $\phi$.}
\begin{equation}
\Delta_Z=\Delta_\gamma\hskip 1pt\frac{1}{v\hskip 1pt\chi}\hskip 1pt 
\frac{(a^\prime/a)}{(a^\prime/a)-(v^\prime/v)}.\label{relation}\end{equation} 
Such relation is rather unique, and does not depend on either $\phi$ or 
$M_{Z_2}$, but only on ratios of the fermionic couplings. In the case 
considered here, $v^\prime$ and $a^\prime$ are explicitly parametrized in 
terms of a (generally unconstrained) angle $\beta$ which characterizes the 
direction of the $Z^\prime$-related extra $U(1)$ generator in the $E_6$ group 
space, and reflects the pattern of symmetry breaking to $SU(2)_L\times U(1)_Y$ 
\cite{rizzo, zwirner}: 
\begin{equation}
v^\prime=\frac{\cos\beta}{c_W\hskip 1pt\sqrt6};\qquad\quad
a^\prime=\frac{1}{2\hskip 1pt c_W\sqrt 6}\left(\cos\beta+
\sqrt{\frac{5}{3}}\sin\beta\right).\label{vprime}\end{equation}
\par 
In Fig.~2 we depict, as an illustration, the cases corresponding to the models 
currently called $\chi$, $\psi$ and $\eta$, corresponding to the choices 
$\beta=0;\hskip 2pt \pi/2;\hskip 2pt \pi-\arctan\sqrt{5/3}\approx 128^\circ$,
respectively. From this figure, two main 
conclusions can be drawn: {\it i}) polarization systematically reduces the 
allowed range for the model parameters, in some cases ($\psi$, $\eta$ models) 
more spectacularly than in other ones ($\chi$ model), and {\it ii}) 
depending on the considered model, different polarization values are relevant, 
i.e., for the $\eta$ and $\psi$ cases $\sigma^R$ is essential while 
for the $\chi$ model $\sigma^L$ provides the main constraint. 
\par 
As a simple quantitative illustration of these features, we can consider 
in more detail the case of the $\eta$ model. As it can be read from Fig.~2,  
for this case the bounds obtainable from $\sigma^{unpol}$ only are the 
following: 
\begin{equation}\begin{array}{l}
-0.005<\Delta_\gamma<0.005 \\
-0.003<\Delta_Z<0.003.\end{array}\label{del2}\end{equation}
\par\noindent 
The use of $\sigma^R$ allows the improvement of Eq.~(\ref{del2}) to the more 
stringent bounds: 
\begin{equation}\begin{array}{l}
-0.002<\Delta_\gamma<0.002 \\
-0.001<\Delta_Z<0.001.\end{array}\label{del3}\end{equation}
\par 
The ranges of $\Delta_\gamma$ and $\Delta_Z$ allowed to the specific models
in Fig.~2 can be translated into limits on the mixing angle $\phi$ and the
heavier gauge boson mass $M_{Z_2}$, using
Eqs.~(\ref{ddeltag})-(\ref{relation}). Continuing our illustrative example of 
the $\eta$ model, in this case the resulting allowed region (at the 95\% CL) 
in the ($\phi,M_{Z_2}$) plane is limited by the thick solid line in Fig.~3. We 
have chosen for $\Delta M$ in Eq.~(\ref{chi}) an upper limit of 
about $150-200\hskip 1pt MeV$, although the 
limiting curves do not appreciably depend on the specific value of this 
quantity. Also, the indicative current lower bound on $M_{Z_2}$ from direct 
searches \cite{pdg}, as well as the bound obtainable from full exploitation of 
the $e^+e^-$ annihilation into lepton-antilepton pairs with polarized 
electrons \cite{riemann,bardin}, 
are reported in Fig.~3. For a comparison, the dashed line in Fig.~3 represents 
the maximal region allowed to $\phi$ by the model-and process-independent 
relation (\ref{phi}), with the same upper bound on $\Delta M$. Finally, the 
thin solid lines exemplify the allowed region for 
a particular $E_6$ `superstring inspired' model with all Higgses belonging 
to the 27-plet representation, such that the mixing angle can be related 
to the masses of $Z_1$ and $Z_2$ as \cite{rizzo}:  
\begin{equation} \phi\simeq {\cal C}\frac{M_{Z_1}^2}{M_{Z_2}^2}.
\label{superstring}\end{equation}
In this case, with $\sigma$ the ratio of appropriate Higgs vacuum expectation 
values squared: 
\begin{equation}{\cal C}=4s_W\left(\frac{\cos\beta}{2\sqrt 6}
-\frac{\sigma-1}{\sigma+1}\frac{\sqrt10\sin\beta}{12}\right).
\label{higgses}\end{equation} 
As one can conclude from Fig.~3, the $W^+W^-$ channel with 
polarized electron beams represents a quite sensitive, and independent, source 
of information on deviations from the SM model due to the extra $Z$, which 
can be combined with that provided by the final leptonic channel and nicely 
complements it. 
\par
One may observe that, in principle, the deviations $\Delta_\gamma$ 
and $\Delta_Z$ in Eqs.~(\ref{ddeltag}) and (\ref{ddeltaz}) are 
energy-dependent, reflecting the energy dependence of (\ref{coupl1}) and 
(\ref{coupl2}) through the neutral vector boson propagators. 
Numerically, however, for the considered values of the energy 
$\sqrt s=0.5\hskip 1pt TeV$ and $M_{Z^\prime}$ such that 
$M_Z\ll\sqrt s\ll M_{Z^\prime}$, one expects $\Delta_\gamma$ and $\Delta_Z$ 
to be dominated by the contributions $\Delta_\gamma (Z_1)$ and 
$\Delta_Z (Z_1)$. Indeed, $\Delta_\gamma (Z_2)$ and $\Delta_Z(Z_2)$ should be 
suppressed relatively to $\Delta_\gamma(Z_1)$ and $\Delta_Z(Z_1)$ by the 
ratio $\vert\chi_2/\chi\vert$, which is of the order of $3\times 10^{-1}$ and 
$3\times 10^{-2}$ for ${M_{Z_2}=1\hskip 1pt TeV}$ and ${3\hskip 1pt TeV}$, 
respectively. For these 
values of $\sqrt s$ and $M_{Z^\prime}$, $\Delta_\gamma (Z_1)$ and 
$\Delta_Z(Z_1)$ are almost energy-independent and therefore so are 
$\Delta_\gamma$ and $\Delta_Z$. Correspondingly, in this case, if we 
define the sensitivity of process (\ref{proc}) to the deviations from the SM 
by the statistical significance
\begin{equation}
{\cal S}=\frac{\sigma-\sigma^{SM}}{\delta\sigma}=\frac{\Delta\sigma}
{\sqrt{\sigma^{SM}}}\sqrt{L_{int}},\label{sensitivity}\end{equation}
where $\delta\sigma$ is the statistical uncertainty and $L_{int}$ the 
time-integrated luminosity, such $\cal S$ is determined only by the explicit 
$s$-dependence of Eq.~(\ref{amplis2}) and is found to behave, with 
$\sqrt s$, as ${\cal S}\propto\sqrt{L_{int}s}$ at fixed $L_{int}$. 
Conversely, for lighter $Z^\prime$, the $Z_2$ contribution can become more 
significant and somewhat modify the $s$-behaviour of the sensitivity. 
\par
The previous discussion should have shown, hopefully in a clear and simple way, 
the advantages of longitudinal electron polarization at the LC in order to 
study the effects of a general model with an extra $Z$ in 
the reaction (\ref{proc}) at a linear electron-positron collider. 
Actually, our analysis has focused on the derivation of bounds, starting 
from the (negative) assumption that no effects, i.e., no deviations from the SM 
predictions are observed within the expected accuracy on the cross section. 
In the next section we shall take, instead, the (positive) attitude of 
assuming that certain deviations from the SM are observed in $\sigma^L$ and/or  
$\sigma^R$. In such a case, it might be possible to identify, to some 
extent, the relevant source of the observed deviation.

\section{Comparison with a model with anomalous gauge couplings}
It has already been pointed out \cite{verze3,pankov} that a model with one 
extra $Z$ would produce virtual manifestations in the final $W^+W^-$ channel 
at the LC that in principle could mimic those of a model (of completely 
different origin) with anomalous trilinear gauge boson couplings. As shown by 
Eqs.~(\ref{amplis2})-(\ref{coupl2}), this is due to the fact that the effects 
of the extra $Z$ can be reabsorbed into a redefinition of the $VWW$ couplings 
($V=\gamma,\hskip 2pt Z$). Therefore, the identification of such an 
effect, if observed at the LC, becomes a relevant problem. 
\par
Using the notations of, e.g., Ref.~\cite{gounaris}, the relevant trilinear 
$WWV$ interaction which conserves $U(1)_{e.m.}$, C and P, can be written as 
($e=\sqrt{4\pi\alpha_{em}}$):
\begin{eqnarray}{\cal L}_{eff}&=&-ie(1+\delta_\gamma)
\left[A_\mu\left(W^{-\mu\nu}W^+_\nu-
W^{+\mu\nu}W^-_\nu\right)+F_{\mu\nu}W^{+\mu}W^{-\nu}\right]
\nonumber\\
&-& ie\hskip 2pt \left(\cot\theta_W+\delta_Z\right)\hskip 2pt \left[Z_\mu
\left(W^{-\mu\nu}W^+_\nu-W^{+\mu\nu}W^-_\nu\right)+Z_{\mu\nu}W^{+\mu}W^{-\nu}
\right]\nonumber\\
&-&ie\hskip 2pt
x_{\gamma}\hskip 2pt F_{\mu\nu}W^{+\mu}W^{-\nu}-
ie\hskip 2pt x_Z\hskip 2pt Z_{\mu\nu}W^{+\mu}W^{-\nu}\nonumber\\
&+&ie\hskip 2pt\frac{y_{\gamma}}{M_W^2}\hskip 2pt 
F^{\nu\lambda}W^-_{\lambda\mu}
W^{+\mu}_{\ \ \nu}+ie\hskip 2pt\frac{y_Z}{M_W^2}\hskip 2pt
Z^{\nu\lambda}
W^-_{\lambda\mu}W^{+\mu}_{\ \ \nu}\hskip 2pt ,\label{lagra}\end{eqnarray}
where $W_{\mu\nu}^{\pm}=\partial_{\mu}W_{\nu}^{\pm}-
\partial_{\nu}W_{\mu}^{\pm}$ and ${Z_{\mu\nu}=\partial_{\mu}Z_{\nu}-
\partial_{\nu}Z_{\mu}}$. In the SM at the tree-level, the anomalous couplings 
in (\ref{lagra}) vanish: $\delta_\gamma=\delta_Z=x_\gamma=x_Z=y_\gamma=y_Z=0$.
\par 
Corresponding to (\ref{lagra}), the helicity amplitudes ${\cal M}^{(\lambda)}$ 
have the structure shown in Tab.~1 \cite{renard}, where $\tau$ and 
$\tau^\prime$ indicate the various possible $W^+$ and $W^-$ polarizations. 
\begin{table}
\centering
\caption{Helicity amplitudes for $e^+e^-\to W^+W^-$}
\begin{tabular}{|c|c|c|}
\hline
$e^+_{-\lambda} e^-_\lambda\to W^+_LW^-_L$  & 
$ \tau=\tau^\prime=0$&$$\\ 
$$&$-\frac{e^2 S\lambda}{2}\sin\theta$ & $$\\ \hline
$\frac{2\lambda-1}{4\hskip 2pt t\hskip 2pt s^2_W}$ &
$\frac{S}{2M_W^2}[\cos\theta-\beta_W (1+\frac{2M_W^2}{S})]$ &\\
\hline
$-\frac{2(1+\delta_\gamma)}{S}+
\frac{2(\cot\theta_W+\delta_Z)}{S-M_Z^2}(v-2a\lambda)$ &
$-\beta_W(1+\frac{S}{2M^2_W})$ &\\ \hline
$ -\frac{x_\gamma}{S}+
\frac{x_Z}{S-M_Z^2}(v-2a\lambda)$ &
$-\beta_W\frac{S}{M^2_W}$ &\\ \hline
\hline
$e^+_{-\lambda} e^-_\lambda\to W^+_TW^-_T$ & $\tau=\tau^\prime=
\pm 1$ & $\tau=-\tau^\prime=\pm 1$ \\
& $-\frac{e^2 S\lambda}{2}\sin\theta$
& $-\frac{e^2 S\lambda}{2}\sin\theta$\\
\hline
$\frac{2\lambda-1}{4\hskip 2pt t\hskip 2pt s^2_W}$ &
$\cos\theta-\beta_W $ & $-\cos\theta-2\tau\lambda $\\
\hline
$-\frac{2(1+\delta_\gamma)}{S}+\frac{2(\cot\theta_W+\delta_Z)}
{S-M_Z^2}(v-2a\lambda)$ &
$-\beta_W $ & $0$ \\ \hline
$-\frac{y_\gamma}{S}+
\frac{y_Z}{S-M_Z^2}(v-2a\lambda)$ &
$-\beta_W\frac{S}{M^2_W}$ & $0$ \\ \hline
\hline
$e^+_{-\lambda} e^-_\lambda\to W^+_TW^-_L$ &
$\tau=0$, $\tau^\prime=\pm 1$ &
$\tau=\pm 1$, $\tau^\prime=0$ \\
& $-\frac{e^2 S\lambda}{2\sqrt{2}}(\tau^\prime\cos\theta-2\lambda)$
& $\frac{e^2 S\lambda}{2\sqrt{2}}(\tau\cos\theta+2\lambda)$\\
\hline
$\frac{2\lambda-1}{4\hskip 2pt t\hskip 2pt S^2_W}$ &
$\frac{\sqrt{S}}{2M_W}[\cos\theta(1+\beta_W^2)-2\beta_W]- $ &
$\frac{\sqrt{S}}{2M_W}[\cos\theta(1+\beta_W^2)-2\beta_W]-$\\
& $-\frac{2M_W}{\sqrt{S}}\frac{\tau^\prime\sin^2\theta}
{\tau^\prime\cos\theta-2\lambda}$ &
$-\frac{2M_W}{\sqrt{S}}\frac{\tau\sin^2\theta}
{\tau\cos\theta+2\lambda}$ \\
\hline
$-\frac{2(1+\delta_\gamma)}{S}+\frac{2(\cot\theta_W+\delta_Z)}
{S-M_Z^2}(v-2a\lambda)$ &
$-\beta_W\frac{\sqrt{S}}{M_W}$ &
$-\beta_W\frac{\sqrt S}{M_W}$ \\
\hline
$-\frac{x_\gamma+y_\gamma}{S}+
\frac{x_Z+y_Z}{S-M_Z^2}(v-2a\lambda)$ &
$-\beta_W\frac{\sqrt S}{M_W}$ &
$-\beta_W\frac{\sqrt S}{M_W}$ \\
\hline
\end{tabular}
\label{tab:tab1}
\end{table}
Here, for practical purposes, the Yang-Mills parts and their deviations 
proportional to $\delta_\gamma$ and $\delta_Z=g_{WWZ}-cot\theta_W$ are 
reported separately from the anomalous `magnetic' and `quadrupole' terms  
conventionally denoted as, respectively  
\begin{equation}
\Delta k_\gamma=k_\gamma-1=x_\gamma;\quad \Delta k_Z\hskip 2pt g_{WWZ}
=(k_Z-1)g_{WWZ}=x_Z,\label{magnetic}\end{equation}
and 
\begin{equation} 
\lambda_\gamma=y_\gamma;\qquad \lambda_Z\cot\theta_W=y_Z.
\label{quadrupole}\end{equation}
The deviations $\Delta_\gamma$ and $\Delta_Z$ needed in 
Eqs.~(\ref{amplis2})-(\ref{deltaz}) are easily derived from the relevant 
entries in Tab.~1. Differently from the previous case of the $Z^\prime$, 
where $\Delta_\gamma$ and $\Delta_Z$ have an explicit (although numerically not 
quite significant) $s$-dependence through Eqs.~(\ref{coupl1}) and 
(\ref{coupl2}), the anomalous trilinear gauge boson couplings are taken 
here as effective {\it constants}. As a consequence, one must assume 
$\delta_\gamma\equiv 0$ to ensure $U(1)_{e.m.}$ gauge invariance. As a matter 
of fact, for maximum generality one might allow for some $s$-dependence 
of the anomalous couplings in (\ref{lagra}) {\it via} form factors 
(in particular, the one relevant to $\delta_\gamma$ must vanish at $s=0$).  
Clearly, that would complicate a true model-independent analysis by the 
introduction of an overwhelming number of independent parameters.
\par
According to the present viewpoint, anomalous gauge couplings are 
understood to parametrize some New Physics defined at a scale $\Lambda$ much 
greater than the Fermi scale, 
$\Lambda\gg v=(\sqrt 2 G_F)^{-1/2}\simeq 250\hskip 1pt GeV$, and involving new, 
very heavy, particles. After integration of the heavy degrees of freedom, a 
`residual' low-energy interaction among the standard `light' degrees of 
freedom should remain. It is natural to assume that also the new physics 
respects $SU(2)\times U(1)$ gauge symmetry spontaneously broken by the Higgs 
vacuum expectation value. The weak interaction is then described by an 
effective Lagrangian, representing a `low-energy' expansion in 
powers of the small ratio $v^2/\Lambda^2$ (and $s/\Lambda^2$) \cite{hagiwara}:
\begin{equation}{\cal L}_W={\cal L}_{SM}+\sum_{d\geq 6}\sum_k
\frac{f^{(d)}_k}{\Lambda^{d-4}} O^{(d)}_k.
\label{lagra1}\end{equation}
The second term in the RHS of Eq.~(\ref{lagra1}) contains the anomalous 
trilinear gauge boson couplings, and is given by $SU(2)\times U(1)$ gauge 
invariant operators $O^{(d)}_k$ made of $\gamma$, $W$, $Z$ and Higgs fields, 
with dimension $d$. The values of the corresponding coupling constants 
$f_k^{(d)}$ are not fixed by the symmetry and therefore must be considered as 
{\it a priori} arbitrary constants, to be determined from experimental data 
(for a detailed analysis and the explicit 
expressions of the relevant operators we refer to \cite{gounaris1}). Clearly, 
in this framework, the lower dimension operators in (\ref{lagra1}) are 
expected to be the leading ones, the higher ones being suppressed by inverse 
powers of the large scale $\Lambda$.\footnote{One can notice that in 
Eq.~(\ref{lagra}) $\delta_V$ and $x_V$ multiply dimension 4 operators, 
while $y_V$ multiply dimension 6 operators. Thus, more consistently with 
(\ref{lagra1}), the latter ones should be scaled to $\Lambda^2$ rather than 
$M^2_W$ as it is usually done.}  
Therefore, while Eq.~(\ref{lagra1}) potentially includes all the anomalous 
{\it constants} of Eq.~(\ref{lagra}) as well as their possible slopes in $s$, 
in practice such slopes are generated by the higher dimension ($d\geq 8$) 
operators and, due to the suppression, are assumed to give a negligible 
effect in the numerical analysis. Specifically, the slope $\delta^\prime$ 
of the $W^+W^-\gamma$ Yang-Mills coupling, 
$\delta_\gamma=s\hskip 1pt \delta^\prime$, is generated by a $d=8$ operator, 
while the other slopes involve $d\geq 10$ operators 
\cite{gounaris1}. Furthermore, it can be shown that the assumption of 
`custodial' global $SU(2)$ symmetry of the New Physics, which naturally 
accounts for the smallness of the $\Delta\rho$ parameter, would imply 
$\delta^\prime=0$ at the $d=8$ level, because the relevant operator would not 
respect this symmetry. 
\par 
It might be useful to recall that the truncation of the sum in 
Eq.~(\ref{lagra1}) to the lowest significant dimension $d=6$ would allow the 
reduction of the number of independent operators, and the corresponding 
anomalous coupling constants, to three (and all slopes identically vanishing) 
\cite{hagiwara,gounaris1}. In this case, by choosing, e.g., as independent 
couplings $\delta_Z$, $x_\gamma$ and $y_\gamma$, one has the relation
\begin{equation}
x_Z=-x_\gamma\tan\theta_W;\qquad y_Z=y_\gamma\cot\theta_W.
\label{xgamma}\end{equation}
Additional assumptions, or specific dynamical models, allow to further reduce 
the number of independent anomalous constants (see, e.g., \cite{gounaris}). 
\par 
As far as the present information on the five anomalous couplings 
in (\ref{lagra}) is concerned, 
{\it indirect} constraints on $WW\gamma$ and $WWZ$ vertices have been obtained 
by comparing low-energy data ($\sqrt s<2\hskip 1pt M_W$) with SM predictions 
for observables that can involve such vertices at the loop level, therefore 
could affect the electroweak corrections. 
The results are \cite{burgess}: ${\delta_Z=-0.059\pm 0.056}$, 
${\Delta k_\gamma=0.056\pm 0.056}$, ${\Delta k_Z=-0.0019\pm 0.0440}$, 
${\lambda_\gamma=-0.036\pm 0.034}$ and ${\lambda_Z= 0.049\pm 0.045}$. 
These constraints are obtained from a global analysis of the data by taking 
the trilinear couplings independently one by one, and fixing the remaining 
ones at the SM values. However, allowing the simultaneous presence of all 
five trilinear anomalous couplings in a multiparameter fit, due to the 
possibility of cancellation and/or correlations, the limits obtained from 
such analysis would considerably weaken to about ${\cal O}(0.1-1)$ or so. 
\par 
For our analysis, we would have now to account for the deviations from 
the SM induced by the various anomalous couplings on polarized observables, 
considering that the general model in Eq.~(\ref{lagra}) 
introduces five independent parameters. Thus, regardless of the attempt to 
distinguish this case from the simpler one of an extra $Z$ (where only two 
parameters are involved), the determination of suitable experimental 
observables, depending on reduced subsets of anomalous gauge boson couplings, 
would represent an important issue by itself which deserves a separate 
treatment. 
\par 
To illustrate a `minimal' identification program, as anticipated in the 
previous section, we assume that a virtual signal has been detected to a given, 
conventionally fixed confidence level, in either $\sigma^L$ or $\sigma^R$, or 
both. In our notations, that would be expressed as: 
\begin{equation} 
\frac{\Delta\sigma^{L,R}}{\delta\sigma^{L,R}}
=\frac{\sigma^{L,R}_{exp}-\sigma^{L,R}_{SM}}{\delta\sigma^{L,R}_{SM}}
\geq \kappa,\label{delsig}\end{equation}
where $\delta\sigma$ is the expected statistical uncertainty on the cross 
section and the value of $\kappa$ corresponds to an assigned number of 
standard deviations. 
\par
As the next step, we try to define an observable which is `orthogonal' to the 
$Z^\prime$ model, in the sense that such variable should depend only on 
those four couplings ($x_V,y_V$) that are specific of the Lagrangian 
(\ref{lagra}), but not on $\delta_Z$ which would induce an effect in 
common with the $Z^\prime$ model of Sect.~2. 
\par 
An illustrative, simple, example of such quantity can be worked out by 
introducing the following polarized observables, along the lines proposed in 
\cite{renard}:
\begin{equation} \sigma{\cal A}^-_{FB}=
\int\limits_{0}^{1}\frac{d\sigma^-}{d\cos\theta}d\cos\theta
-\int\limits_{-1}^{0}\frac{d\sigma^-}{d\cos\theta}d\cos\theta, 
\label{afb}\end{equation}
and
\begin{equation} \sigma {\cal A}^\pm_{CE}(z^*)=
\int\limits_{-z^*}^{z^*}\frac{d\sigma^\pm}{d\cos\theta}d\cos\theta-
\left[\hskip 2pt\int\limits_{z^*}^{1}\frac{d\sigma^\pm}{d\cos\theta}d\cos\theta+
\int\limits_{-1}^{-z^*}\frac{d\sigma^\pm}{d\cos\theta}d\cos\theta\right]. 
\label{acep}\end{equation}
Similar to (\ref{xsection}), one can expand (\ref{afb}) and (\ref{acep}) as
\begin{equation}\sigma{\cal A}^-_{FB}=\frac{\pi^2\alpha_{em}^2\beta_W}
{2s}\left[F_0^-{\cal O}_{0,FB}+F_2^-{\cal O}_{2,FB}
+\left(F_6^-+F_7^-\right){\cal O}_{6,FB}\right],\label{aeffebi}
\end{equation}
\begin{equation}\sigma{\cal A}^\pm_{CE}(z^*)=
\frac{\pi\alpha_{em}\beta_W}{2s}\sum_i F^\pm_i{\cal O}_{i,CE}(z^*),
\label{acie}\end{equation}
where the explicit expressions of couplings and corresponding kinematical 
coefficients can be obtained from the Appendix. One can notice, also, that 
$\sigma{\cal A}^+_{FB}=0$. The $\delta_Z$ contribution in Eqs.~(\ref{aeffebi}) 
and (\ref{acie}) is contained in $F_1^\pm$ and $F_2^-$ (recall that 
${F_2^+=0}$). Therefore, the simplest observable `orthogonal' to $\delta_Z$ is 
represented, at $\sqrt s=500\hskip 2pt GeV$, by the quantity
\begin{equation} \sigma{\cal A}^+_{CE}(z^*\simeq 0.4).\label{aceplus}
\end{equation}
Indeed, numerical inspection of the formulae in the Appendix shows that, 
at $z^*\simeq 0.4$ (and $\sqrt s=500\hskip 2pt GeV$) the coefficient 
${\cal O}_{1,CE}$ vanishes, leaving a pure dependence of (\ref{aceplus}) 
from $(x_V^+,y_V^+)$ but not from $\delta_Z$. Clearly, the position of this 
zero is entirely determined by $M_W$ and the CM energy $\sqrt s$. The 
possibility to measure (\ref{aceplus}) strongly depends on the angular 
resolution for the $W$ and its decay products.\par
Another example, which would eliminate both $F_1^-$ and $F_2^-$ in 
Eqs.~(\ref{aeffebi}) and (\ref{acie}), should be the specific combination:
\begin{equation}Q^-=\sigma{\cal A}^-_{FB}-
\frac{{\cal O}_{2,FB}}{{\cal O}_{2,CE}(0.4)}\hskip 2pt 
\left(\sigma{\cal A}_{CE}^-(0.4)\right)\simeq \sigma{\cal A}^-_{FB}
+0.029\cdot\sigma{\cal A}_{CE}^-(0.4).\label{qminus}\end{equation}
\par 
Still another possibility could be represented, in principle, by 
the combination 
\begin{equation}
P^+(z^*)=\sigma^+-\frac{{\cal O}_1}{{\cal O}_{1,CE}(z^*)}\left(
\sigma{\cal A}^+_{CE}(z^*)\right),\label{pplus}\end{equation}
with $\sigma^+$ the total cross section for right-handed electrons and 
arbitrary $z^*\not\simeq 0.4$. The coefficients 
${\cal O}_1$, ${\cal O}_{1,CE}$ and  
${\cal O}_{2,CE}$ can be easily calculated from the formulae given in the 
Appendix. Concerning the dependence 
on the remaining anomalous couplings, in the linear approximation to the 
$F^\pm_i$ (see Eq.~(A4)) the combination (\ref{qminus}) depends on 
$(x_V^-,y_V^-)$, while (\ref{pplus}) is determined by $(x_V^+,y_V^+)$ 
similar to (\ref{aceplus}), but with different parametrical dependence 
hence with different sensitivities on the anomalous couplings.\par
In Figs.~4-6 we depict the statistical significance of the variables 
(\ref{aceplus})-(\ref{pplus}) as a function of the relevant anomalous 
couplings. As anticipated in the previous section, for any observable $O$ 
such significance is defined as the ratio 
${\cal S}=\Delta O/(\delta O)_{stat}$ where $\Delta O=O(x_V,y_V)-
O^{SM}$ and $(\delta O)_{stat}$ is the statistical uncertainty attainable on 
$O$. The different sensitivities to the various couplings $x_V$ and $y_V$ 
can be directly read from these figures and, as one can see, in certain cases 
they can be substantial. Figs.~4-6 are obtained by varying one of the 
parameters at a time and setting all the other ones at the SM values.\par
By definition, the observables (\ref{aceplus})-(\ref{pplus}) require 
100\% longitudinal electron polarization, a situation that will not be fully 
obtained in practice. However, the presently planned degree of electron 
polarization at the NLC, $\vert P_L\vert\simeq 0.90-0.95$ 
\cite{prescott,settles} is 
high enough that Eq.~(\ref{longi}) can represent a satisfactory approximation 
to evaluate such observables. Clearly, such approximation would be 
substantially improved if also positron polarization were available, e.g., 
$\vert {\bar P}_L\vert\simeq 0.6$ \cite{settles}, because in that case in 
Eq.~(\ref{longi}) the coefficient of $\sigma^+$ could be emphasized over that 
of $\sigma^-$ by a large factor, or viceversa. 
Moreover, we can remark that the independence of $Q$ in Eq.~(\ref{qminus}) 
from $\delta_Z$ holds for both $\pm$ cases and, consequently, also for the 
case of unpolarized beams.
\par
Concerning a possible discrimination between the $Z^\prime$ model of Sect.~2 
and the model considered in this section, a strategy could be the following. 
If a signal is observed in either $\sigma^L$ and/or $\sigma^R$ and also in 
at least one of the `orthogonal' observables defined above, 
we can conclude that it is due to 
the model with anomalous gauge couplings, and we can try to derive the values 
of some of them by properly analyzing the observed effects 
\cite{andreev,likhoded}. 
If, conversely, only $\sigma^L$ and/or $\sigma^R$ show an effect, 
we are left with the possibility that both models are responsible for such 
deviations. In this situation, we still have a simple tool to try to 
distinguish among the two models, which uses the observation that, under the 
assumption that only $\delta_V$ and $\Delta_V$ are effective, the 
expressions of the consequent deviations of the integrated cross sections 
$\sigma^L$ and $\sigma^R$ are, respectively:
\begin{equation}
\Delta\sigma^{R,L}\simeq\Delta\sigma^\pm\propto\delta_\gamma-
\delta_Zg_e^{R,L}\chi,
\label{delsigl}\end{equation}
and 
\begin{equation}
\Delta\sigma^{R,L}\simeq\Delta\sigma^\pm\propto\Delta_\gamma-
\Delta_Zg_e^{R,L}\chi.
\label{delsigr}\end{equation}
Here, both $\Delta_V$ and $\delta_V$ have been taken nonvanishing, and 
$g_e^{L,R}=v\pm a$ are the left- and right-handed electron couplings, 
respectively. However, recalling that $\delta_\gamma=0$ in the case of 
anomalous trilinear gauge boson couplings, using the experimental value of 
$s_W^2\simeq 0.23$, one has for such a model the very characteristic feature
\begin{equation}
\Delta\sigma^L\simeq\left(1-\frac{1}{2s_W^2}\right)\Delta\sigma^R
=-1.17\Delta\sigma^R, \label{feature}
\end{equation}
where the explicit expressions of $g_e^L$ and $g_e^R$ have been used. 
If, on the contrary, 
the effect is due to a model with a $Z^\prime$, no {\it a priori} relationship 
exists between $\Delta\sigma^L$ and $\Delta\sigma^R$. Accordingly, from 
inspection of these two quantities, if they are found not to be related by 
Eq.~(\ref{feature}) to a given confidence level, one would conclude that the 
observed effect should be due to the general extra $Z$ discussed in Sect.~2. 
Then, depending on the actual values of the experimental deviations, a 
determination of the two parameters $\Delta_\gamma$ and $\Delta_Z$ might be 
carried on.
\par
Actually, if the deviations of $\sigma^{L,R}$ satisfy the correlation 
Eq.~(\ref{feature}), a small residual ambiguity would remain. Although the 
possibility that in a model with both $\Delta_\gamma$ and $\Delta_Z$ 
nonvanishing the correlation Eq.~(\ref{feature}) is satisfied just by chance 
seems rather unlikely, one cannot exclude it {\it a priori}. Should this be 
the real situation, further analysis, e.g., in the different final 
fermion-antifermion channel would be required. The discussion of this 
essentially unlikely case can be performed, but is beyond the purpose of 
this paper.

\section{Concluding remarks}
We have shown in this paper that the availability of longitudinal electron 
beam polarization at the LC would be very useful for the study of the most 
general model with one extra $Z$ from an analysis of the final $W^+W^-$ 
channel. In principle, it would also be possible to discriminate this model 
from a rather `natural' competitor one where anomalous gauge boson couplings 
are present. This could be done by analyzing suitable experimental 
variables, all defined in the same $W^+W^-$ final channel.
\par
The interesting property of polarized observables in the $W^+W^-$ channel 
should be joined to analogous interesting features that are characteristic of 
polarization asymmetries in the final two-fermion channel, whose general 
discussion has been presented recently \cite{verze4}.
\par 
All these facts allow us to conclude that polarization at the LC would be, 
least to say, a highly desirable opportunity.
 
\vspace{2.0cm}

\section*{Acknowledgements}
\noindent
AAP acknowledges the support and the hospitality of INFN-Sezione di 
Trieste and of the International Centre for Theoretical Physics, Trieste.\par
\noindent NP acknowledges the support and hospitality of the Laboratorio de 
F\'isica Te\'orica, Departamento de F\'isica, Universidad Nacional de La Plata 
(Argentina), where part of this work was done. He is grateful to 
Prof. C. Garcia Canal for his kind invitation and numerous stimulating 
discussions. 

\vfill
\eject
\newpage

\section*{Appendix}
Limiting to CP conserving couplings, to generally describe the deviations from 
the SM of the cross section for process (\ref{proc}) of interest 
here, we have to account for the effect of six parameters, i.e., the five 
anomalous gauge couplings ($\delta_Z$, $x_V$ and $y_V$) of Eq.~(\ref{lagra}) 
plus the deviation $\delta_\gamma$ possible in the $Z^\prime$ model. Moreover, 
in this case the index $i$ in Eq.~(\ref{xsection}) runs from 1 to 
11.\footnote{The general expansion including also 
CP violating couplings can be found, e.g., in \cite{renard}.} 
\par 
Referring to the expression of polarized differential cross sections 
$d\sigma^\pm$ in Eq.~(\ref{xsection}), one can easily realize that 
the $F_i$ there are conveniently expressed in terms of the combinations of 
anomalous coupling constants defined as follows: 
$$\delta_V^\pm=\delta_\gamma-\delta_Z\hskip 2pt g_e^\pm\hskip 2pt\chi
\hskip 2pt;\quad 
x_V^\pm=x_\gamma-x_Z\hskip 2ptg_e^\pm\hskip 2pt\chi\hskip 2pt;\quad 
y_V^\pm=y_\gamma-y_Z\hskip 2ptg_e^\pm\hskip 2pt\chi\hskip 2pt,\eqno(A1)$$
where
$$g_e^+=v-a=\tan\theta_W\hskip 2pt;
\qquad g_e^-=v+a=g_e^+\left(1-\frac{1}{2s_W^2}\right), \eqno(A2)$$ 
and $v$, $a$ and the $Z$ propagator $\chi$ have been previously defined in 
Sect.~2 with regard to Eqs.~(\ref{amplis}) and (\ref{amplis1}). Introducing 
the combination
$$g_s^\pm=1-\cot\theta_W\hskip 2ptg_e^\pm\hskip 2pt\chi, \eqno(A3)$$
we have:
$$ F_0^-=\frac{1}{16s_W^4}\hskip 2pt;\quad 
F_1^\pm=\left(g_s^\pm+\delta_V^\pm\right)^2\hskip 2pt;\quad 
F_2^-=-\frac{1}{2s_W^2}\left(g_s^-+\delta_V^-\right)\hskip 2pt,$$ 
$$ F_3^\pm=\left(g_s^\pm+\delta_V^\pm\right)\hskip 1pt x_V^\pm
\simeq g_s^\pm\hskip 1pt x_V^\pm\hskip 2pt;\quad 
F_4^\pm=\left(g_s^\pm+\delta_V^\pm\right)\hskip 1pt y_V^\pm\simeq 
g_s^\pm\hskip 1pt y_V^\pm, $$
$$F_6^-=-\frac{1}{4s_W^2}\hskip 1pt x_V^-\hskip 2pt;\qquad 
F_7^-=-\frac{1}{4s_W^2}\hskip 1pt y_V^-\hskip 2pt, $$
$$F_9^\pm=\frac{1}{2}\left(x_V^\pm\right)^2\hskip 2pt;\quad 
F_{10}^\pm=\frac{1}{2}\left(y_V^\pm\right)^2\hskip 2pt;\quad 
F_{11}^\pm=\frac{1}{2}\left(x_V^\pm\hskip 1pt y_V^\pm\right).\eqno(A4)$$ 
All other $F$'s vanish. Clearly, the polarized cross sections will depend on  
on the anomalous parameters as 
$\sigma^+=\sigma^+(\delta_V^+,x_V^+,y_V^+)$ and 
$\sigma^-=\sigma^-(\delta_V^-,x_V^-,y_V^-)$.
\par   
With ${\displaystyle t=M_W^2-\frac{s}{2}\left(1-\beta_W\cos\theta\right)}$, 
and $\beta_W$ defined previously, 
the corresponding kinematical coefficients ${\cal O}_i(s,\cos\theta)$ 
that appear in Eq.~(\ref{xsection}) are \cite{renard}:
$${\cal O}_0=8\hskip 1pt\left[\frac{2s}{M_W^2}+
\frac{\beta_W^2}{2}\left(\frac{s^2}{t^2}+
\frac{s^2}{4M_W^4}\right)\sin^2\theta\right], $$
$${\cal O}_1=\frac{\beta_W^2}{8}\left[\frac{16s}{M_W^2}+
\left(\frac{s^2}{M_W^4}-\frac{4s}{M_W^2}+12\right)\sin^2\theta\right],$$
$${\cal O}_2=16\left(1+\frac{M_W^2}{t}\right)+8\beta_W^2\left[\frac{s}{M_W^2}+
\frac{1}{16}\left(\frac{s^2}{M_W^4}-\frac{2s}{M_W^2}-
\frac{4s}{t}\right)\sin^2\theta\right],  $$
$${\cal O}_3=\frac{\beta_W^2s^2}{2M_W^4}\left[1+\frac{6M_W^2}{s}-
\left(1-\frac{2M_W^2}{s}\right)\cos^2\theta\right]\hskip 1pt ;
\qquad {\cal O}_4=\frac{4\beta_W^2s}{M_W^2}, $$
$${\cal O}_6=\frac{\beta_Ws^3}{2tM_W^4}\hskip 1pt
\left[-\beta_W\left(1+\frac{6M_W^2}{s}\right)+
\left(1+\frac{4M_W^2}{s}-\frac{16M_W^4}{s^2}\right)\cos\theta \right .$$
$$\left . +\beta_W\left(1-\frac{2M_W^2}{s}\right)\cos^2\theta-
\beta_W^2\cos^3\theta\right], $$
$${\cal O}_7=\frac{4\beta_Ws^2}{tM_W^2}\left[-\beta_W+
\left(1-\frac{2M_W^2}{s}\right)\cos\theta\right],$$
$${\cal O}_9=\frac{\beta_W^2s^2}{2M_W^4}\left[1+\frac{2M_W^2}{s}-
\left(1-\frac{2M_W^2}{s}\right)\cos^2\theta\right], $$
$${\cal O}_{10}=\frac{\beta_W^2s^2}{M_W^4}\left[1+\frac{M_W^2}{s}-
\left(1-\frac{M_W^2}{s}\right)\cos^2\theta\right]\hskip 1pt ;\quad 
{\cal O}_{11}=\frac{2\beta_W^2s}{M_W^2}\left(1+\cos^2\theta\right).
\eqno(A5)$$
\par\noindent  
Then, defining
$$C=\frac{2M_W^2-s}{2p\sqrt s}=-\frac{1+\beta_W^2}{2\beta_W};\quad 
L_{FB}=\ln{\frac{C^2-1}{C^2}}, $$
the coefficients ${\cal O}_{i, FB}$ in Eq.~(\ref{aeffebi}) can be expressed 
as:
$${\cal O}_{0, FB}=32\left[\frac{1}{C}+CL_{FB}\right]\hskip 1pt; \quad 
{\cal O}_{2, FB}=4\left[\beta_W+\frac{4M_W^2}{s\beta_W} 
\left(2+\frac{M_W^2}{s}\right)L_{FB}\right], $$
$${\cal O}_{6, FB}={\cal O}_{7, FB}=32\frac{M_W^2}{s\beta_W}L_{FB}\hskip 1pt .
\eqno(A6)$$
\par\noindent 
Finally, defining 
$$L_{CE}({z^*})=\ln\frac{C+1}{C-1}-2\ln\frac{C+{z^*}}{C-{z^*}}, $$
the coefficients ${\cal O}_{i,CE}({z^*})$ in Eq.~(\ref{acie}) have the 
following expressions:
$${\cal O}_{0,CE}=32\left[\left(2{z^*}-1\right)\left(\frac{s}{M_W^2}-1\right)
+\left({z^*}-\frac{{z^*}^3}{3}-\frac{1}{3}\right)
\frac{\beta_W^2s^2}{8M_W^4}\right .$$
$$\left . 
-2{z^*}\frac{C^2-1}{C^2-{z^*}^2}+1-CL_{CE}({z^*})
\right],$$
$${\cal O}_{1,CE}=\beta_W^2\left[4\left(2{z^*}-1\right)\frac{s}{M_W^2}+
\frac{1}{2}\left({z^*}-\frac{{z^*}^3}{3}-\frac{1}{3}\right)
\left(\frac{s^2}{M_W^4}-\frac{4s}{M_W^2}+12\right)\right],$$
$${\cal O}_{2,CE}=2\left(2{z^*}-1\right)\left(20+\frac{8\beta_W^2s}{M_W^2}-
\frac{8M_W^2}{s}\right)+\frac{2\beta_W^2s^2}{M_W^4}
\left(1-\frac{2M_W^2}{s}\right)\left({z^*}-\frac{{z*}^3}{3}-\frac{1}{3}
\right)$$
$$-16\frac{M_W^2}{s}\left(2+\frac{M_W^2}{s}\right)\frac{1}{\beta_W}
L_{CE}({z^*}), $$
$${\cal O}_{3,CE}=\frac{\beta_W^2s^2}{M_W^4}\left[\left(2{z^*}-1\right)
\left(1+\frac{6M_W^2}{s}\right)+\frac{1}{3}\left(1-2{z^*}^3\right)
\left(1-\frac{2M_W^2}{s}\right)\right],$$
$${\cal O}_{4,CE}=\frac{8\beta_W^2s}{M_W^2}\left(2{z^*}-1\right), $$
$${\cal O}_{6,CE}=\frac{2}{3}\left(1-2{z^*}^3\right)\frac{\beta_W^2s^2}
{M_W^4}+2\left(2{z^*}-1\right)\frac{s^2}{M_W^4}\left(1+\frac{4M_W^2}{s}-
\frac{16M_W^4}{s^2}\right)$$
$$-\frac{32M_W^2}{s}\frac{1}{\beta_W}L_{CE}({z^*}),
$$
$${\cal O}_{7,CE}=\frac{16s}{M_W^2}\left[\left(2{z^*}-1\right)
\left(1-\frac{2M_W^2}{s}\right)-\frac{2M_W^4}{s^2}\frac{1}{\beta_W}L_{CE}
({z^*})\right],$$
$${\cal O}_{9,CE}=\frac{\beta_W^2s^2}{M_W^4}\left[\left(2{z^*}-1\right)
\left(1+\frac{2M_W^2}{s}\right)+\frac{1}{3}\left(1-2{z^*}^3\right)
\left(1-\frac{2M_W^2}{s}\right)\right],$$
$${\cal O}_{10,CE}=\frac{2\beta_W^2s^2}{M_W^4}\left[\left(2{z^*}-1\right)
\left(1+\frac{M_W^2}{s}\right)+\frac{1}{3}\left(1-2{z^*}^3\right)
\left(1-\frac{M_W^2}{s}\right)\right],$$
$${\cal O}_{11,CE}=\frac{8\beta_W^2s}{M_W^2}\left({z^*}+\frac{{z^*}^3}{3}
-\frac{2}{3}\right).\eqno(A7)$$

\newpage
\section*{Figure captions}
\begin{description}

\item{\bf Fig.~1} Allowed bands for $\Delta_\gamma$ and $\Delta_Z$ (95\% CL) 
from $\sigma^L$ and $\sigma^R$ at $\sqrt s=500\hskip 2pt GeV$ and 
$L_{int}=50\hskip 2pt fb^{-1}$, and combined allowed domain. Also the domain 
determined by $\sigma^{unpol}$ alone is reported.

\item{\bf Fig.~2} Same as Fig.~1, with the straight lines (\ref{relation}) 
for the $\eta$, $\chi$ and $\psi$ models superimposed.

\item{\bf Fig.~3} Thick solid contour: allowed domain (95\% CL) in the 
($\phi, M_{Z_2}$) plane for the $\eta$ model from process (\ref{proc}) at 
$\sqrt s=0.5\hskip 1pt TeV$. Dotted lines: constraints from the 
$Z-Z^\prime$ mass-matrix relation (\ref{phi}) with the upper limit 
$\Delta M=200\hskip 1pt MeV$. Thin solid contours: constraints for the 
particular case of Eq.~(\ref{superstring}) with $\sigma=0$ and 
$\sigma=\infty$ in Eq.~(\ref{higgses}). The current limit on $M_2$ from 
direct searches and that expected from $e^+e^-\to l^+l^-$ at 
$\sqrt s=0.5\hskip 1pt TeV$ with polarized electrons are also indicated.

\item{\bf Fig.~4} Statistical significance in $x^+_V$ of the observables  
$\sigma{\cal A}^+_{CE}$ of Eq.~(\ref{aceplus}) and $P^+$ 
of Eq.~(\ref{pplus}) with $z^*=0.5$.

\item{\bf Fig.~5} Statistical significance in $y^+_V$ of the observables 
$\sigma{\cal A}^+_{CE}$ of Eq.~(\ref{aceplus}) and $P^+$ 
of Eq.~(\ref{pplus}) with $z^*=0.5$.

\item{\bf Fig.~6} Statistical significance in $x^-_V,y^-_V$ of the variable 
$Q^-$ of Eq.~(\ref{qminus}). 

\end{description}

\newpage

\begin{thebibliography}{99}

\bibitem{verze1} P. Chiappetta, J. Layssac, F. M. Renard and C. Verzegnassi, 
Phys. Rev. D {\bf 54} (1996) 789; G. Altarelli, N. Di Bartolomeo, F. Feruglio, 
R. Gatto and M. Mangano, Phys. Lett. {\bf B375} (1996) 292. 
\bibitem{lep1} For an updated review see, e.g., A. Blondel, talk at the 
{\it 28th International Conference on High Energy Physics}, 25-31 July 1996, 
Warsaw, Poland. 
\bibitem{sld} For an updated review see, e.g., M. W. Gr\"unewald, talk at the 
{\it 28th International Conference on High Energy Physics}, 25-31 July 1996, 
Warsaw, Poland.  
\bibitem{cdf} E. Buckley-Geer, FERMILAB-Conf-95/316-E (1995), Proceedings of 
the {\it EPS Conference on High Energy Physics}, Brussels, Belgium (1995). 
\bibitem{verze2} G. Montagna, F. Piccinini, J. Layssac, F. M. Renard and 
C. Verzegnassi, Montpellier Report PM 96-25 (1996).
\bibitem{sasha} A. A. Pankov and N. Paver, Phys. Rev. D {\bf 48} (1993) 63.
\bibitem{verze3} P. Chiappetta, F. M. Renard and C. Verzegnassi, Z. Phys. C 
{\bf 71} (1996) 673.
\bibitem{gounaris} See, e.g., G. Gounaris, J. L. Kneur, J. Layssac, 
G. Moultaka, F. M. Renard and D. Schildknecht, Proceedings of the Workshop 
$e^+e^-${\it Collisions at 500 GeV: the Physics Potential}, Ed. P.M. Zerwas 
(1992), DESY 92-123B, p.735.
\bibitem{lep_recent} {\it A Combination of Preliminary Electroweak 
Measurements and Constraints on the Standard Model}, the LEP Collaborations 
ALEPH, DELPHI, L3 and OPAL, report LEPEWWG/97-01.
\bibitem{altarelli} G. Altarelli, CERN report CERN-TH-96-265, presented at the  
{\it NATO Advanced Study Institute on Techniques and Concepts of 
High-Energy Physics}, St. Croix, U.S. Virgin Islands (1996); 
CERN report CERN-TH.7464/94, Proceedings of the Workshop on {\it Radiative 
Corrections: Status and Outlook}, (Gatlinburg, Tenn., USA, 1994), ed. 
B. L. Ward, World Scientific 1995.
\bibitem{babu} K. S. Babu, C. Kolda and J. March-Russell, Phys. Rev. D
{\bf 54} (1996) 4635.
\bibitem{babich} A. A. Babich, A. A. Pankov and N. Paver, Phys. Lett. 
{\bf B346} (1995) 303.
\bibitem{prescott}  C. Y. Prescott, Proceedings of the Workshop {\it Physics
and Experiments with Linear $e^+e^-$ Colliders}, (Waikoloa, Hawaii, 1993),
eds. F. A. Harris, S. L. Olsen, S. Pakvasa and X. Tata, (World Scientific,
Singapore 1993), p. 379.
\bibitem{pankov} A. A. Pankov and N. Paver, Trieste Report IC/96/175, 
UTS-DFT-96-01 (1996).
\bibitem{pdg} Review of Particle Physics, Phys. Rev. D {\bf 54} (1996).
\bibitem{rizzo} For a review see, e.g.: J. L. Hewett, T. G. Rizzo, Physics
Reports {\bf 183} (1989) 193.
\bibitem{zwirner} F. del Aguila, M. Quiros and F. Zwirner, Nucl. Phys. 
{\bf B284} (1987) 530; {\bf B287} (1987) 419. 
\bibitem{riemann} S. Riemann, FORTRAN program ZEFIT Version 4.2.
\bibitem{bardin} D. Bardin {\it et al.}, Report CERN-TH.6443/92 (1992).
\bibitem{renard} G. Gounaris, J. Layssac, G. Moultaka and F. M. Renard, 
Int. J. Mod. Phys. A {\bf 8} (1993) 3285.
\bibitem{hagiwara} See, e.g., K. Hagiwara, S. Ishihara, R. Szalapski and 
D. Zeppenfeld, Phys. Lett. B {\bf 283} (1992) 353; Phys. Rev. D 
{\bf 48} (1993) 2182; and references there.
\bibitem{gounaris1} G. Gounaris and F. M. Renard, Z. Phys. C {\bf 59} (1993) 
133.
\bibitem{burgess} C. P. Burgess, S. Godfrey, H. K\"onig, D. London and 
I. Maksymyk, Phys. Rev. D {\bf 50} (1994) 7011.
\bibitem{settles} P. Zerwas, talk at the ECFA/DESY Linear Collider 
Workshop, Hamburg (Germany), Nov. 20-22, 1996, to appear in the 
Proceedings.
\bibitem{andreev} V. V. Andreev, A. A. Pankov and N. Paver, Phys. Rev. D 
{\bf 53} (1996) 2390.
\bibitem{likhoded} A. A. Likhoded, T. Han and G. Valencia, Phys. Rev. D 
{\bf 53} (1996) 4811. 
\bibitem{verze4} F. M. Renard and C. Verzegnassi, Montpellier Report PM/96-27 
(1996).
\end{thebibliography}
\end{document}